\documentclass[11pt,a4paper]{article}
\usepackage{amsmath}
\usepackage[mathcal]{eucal}
\usepackage{latexsym}
\usepackage{amssymb}
\begin{titlepage}
\title{4-Dimensional General Relativity from the instrinsic spatial geometry of SO(3) Yang--Mills theory}
\author{Eyo Eyo Ita III}
\medskip
\input amssym.def
\input amssym.tex

\def \in{\indent}

\begin{document}
\maketitle
\bigskip  
\centerline{Department of Applied Mathematics and Theoretical Physics} 
\smallskip
\centerline{Centre for Mathematical Sciences, University of Cambridge, Wilberforce Road}
\smallskip
\centerline{Cambridge CB3 0WA, United Kingdom}
\smallskip
\centerline{eei20@cam.ac.uk} 

\bigskip

\begin{abstract}
In this paper we derive 4-dimensional General Relativity from three dimensions, using the intrinsic spatial geometry inherent in Yang--Mills theory which has been exposed by previous authors as well as as some properties of the Ashtekar variables.  We provide various interesting relations, including the fact that General Relativity can be written as a Yang--Mills theory where the antiself-dual Weyl curvature replaces the Yang--Mills coupling constant.  We have generalized the results of some previous authors, covering Einsteins spaces, to include more general spacetime geometries.  
\end{abstract}
\end{titlepage}

\section{Introduction}

\indent
In the Ashtekar formalism one embeds General Relativity into the phase space of a complexified $SO(3)$ Yang--Mills theory.  The Yang--Mills electric field $E^i_a$ is interpreted as densitized triad $\widetilde{\sigma}^i_a$,\footnote{For index conventions internal $SO(3,C)$ indices will be denoted by lowercase symbols from the beginning part of the Latin alphabet $a,b,c,\dots$, and 3-dimensional spatial indices from the middle part $i,j,k\dots$.  Both types of indices take values $1-3$.} given by 
\begin{eqnarray}
\label{YANGMILL111}
\widetilde{\sigma}^i_a={1 \over 2}\epsilon^{ijk}\epsilon_{abc}e^b_je^c_k.
\end{eqnarray}
\noindent
Equation (\ref{YANGMILL111}) is an antisymmetric combination of spatial triads $e^a_i$, while the same set of triads arranged in symmetric combination defines a spatial 3-metric $h_{ij}$, given by
\begin{eqnarray}
\label{YANGMILLS121}
h_{ij}=e^a_ie^a_j.
\end{eqnarray}
\noindent
Equations (\ref{YANGMILL111}) and (\ref{YANGMILLS121}) encode the same information and presumably should lead to alternate but equivalent descriptions of General Relativity.  The Ashtekar formalism, which 
uses $\widetilde{\sigma}^i_a$ as a basic momentum space variable, can be written in 3+1 form as (\cite{ASH1},\cite{ASH2},\cite{ASH3})
\begin{eqnarray}
\label{YANGMILL2}
I_{Ash}=\int{dt}\int_{\Sigma}d^3x\widetilde{\sigma}^i_a\dot{A}^a_i
+A^a_0D_i\widetilde{\sigma}^i_a-\boldsymbol{H}_{Ash}(\widetilde{\sigma},A),
\end{eqnarray}
\noindent
where $A^a_i$ is a $SO(3,C)$ gauge connection with magnetic field $B^i_a$.  The Hamiltonian in (\ref{YANGMILL2}) is given by 
\begin{eqnarray}
\label{YANGMILL3}
\boldsymbol{H}_{Ash}=\epsilon_{ijk}N^i\widetilde{\sigma}^j_aB^k_a+{i \over 2}\underline{N}\epsilon_{ijk}\epsilon^{abc}\widetilde{\sigma}^i_a\widetilde{\sigma}^j_b\Bigl(B^k_c+{\Lambda \over 3}\widetilde{\sigma}^k_c\Bigr),
\end{eqnarray}
\noindent
and the fields $N^{\mu}=(N,N^i)$ are auxilliary fields, respectively the lapse function and shift vector, with $\underline{N}=N(\hbox{det}\widetilde{\sigma})^{-1/2}$.  These fields smear respectively the Hamiltonian and diffeomorphism constraints, and $A^a_0$ in (\ref{YANGMILL2}) smears the Gauss' law constraint.  The action (\ref{YANGMILL2}) is identical in structure with $SO(3)$ Yang--Mills theory, albeit with a different 
Hamiltonian (\ref{YANGMILL3}).\par
\indent
In this paper we will show that General Relativity and Yang--Mills theory are in a sense more literally related.  To accomplish this we will harness the relation of nonabelian gauge theory to intrinsic spatial geometry which has been exposed by previous authors within the purely Yang--Mills context.  Some of the main ideas contained in this paper have been applied in \cite{HIDDENSPATIAL} and \cite{HIDDENSPATIAL1}, where the authors uncover a natural spatial geometry encoded within $SU(2)$ and $SU(3)$ Yang-Mills theory.  It is shown how using locally gauge-invariant quantities, one obtains a geometrization of these gauge theories in the form of a 3-dimensional Einstein space $Q^{(3)}$ with Ricci tensor
\begin{eqnarray}
\label{YANGMILL5}
R_{ij}=kh_{ij}   
\end{eqnarray}
\noindent
where $k$ is a numerical constant.  We would like to generalize this to four spacetime dimensions exhibiting the two local degrees of freedom of General Relativity.  But it is known that 3-dimensional gravity can be described as a topological field theory, with no propagating degrees of freedom \cite{WEETEN}.  We will show that (\ref{YANGMILL5}) actually describes a 4-dimensional geometry if the space $Q^{(3)}$ is allowed to have torsion, which specifically must be identified with the extrinsic curvature of 3-space.\par
\indent
The organization of this paper is as follows.\footnote{The main result of sections 2 through 4 which is new is the association of 3-dimensional torsion with extrinsic curvature.  While some of the background material in these sections might be familiar to the reader and can be skimmed, it is necessary to set the context, framework and notation for sections 5 and 6 which contain the main results of this paper.}  In Section 2 we define a 3-dimensional affine connection $\Gamma^i_{jk}$ and relate its curvature to the Ashtekar magnetic 
field $B^i_a$.  The torsion of $\Gamma^i_{jk}$ has the degrees of freedom of a 3-dimensional extrinsic curvature tensor when Gauss' law holds, which suggests the existence of a 4-dimensional Einstein Hilbert action based upon the space $Q^{(3)}$.  Section 4 puts in place the canonical structure necessary to make this the case, and performs a Legendre map into the Hamiltonian formulation.  In Section 5 we show that $Q^{(3)}$ also defines a Yang--Mills action in curved spacetime, with gravitational degrees of freedom encoded in a field $\Psi_{ae}$ which takes the place of the Yang--Mills coupling constant.  The results of sections 2 and 3 suggest that Einstein's 4-dimensional metric gravity and this Yang--Mills theory are the same theory, provided that the role of the Gauss' constraint can be clarified.  The field $\Psi_{ae}$ takes on the interpretation of the self-dual Weyl tensor, which in Section 6 we show to be the Einstein tensor of $Q^{(3)}$.  The constraints on $\Psi_{ae}$ are a natural consequence of the Bianchi identities for $Q^{(3)}$.  Section 7 is the conclusion and a discussion, where we write down an action principle based on the fields $\Psi_{ae}$ and $A^a_i$.\par  
\indent
In this paper we will make use of three different connections 
$D_i$, $\nabla_i$ and $\overline{\nabla}_i$, which must be defined in order to avoid confusion.  $D_i$ is the the gauge covariant derivative whose action on $SO(3,C)$ 3-vectors $v_a$ is given by
\begin{eqnarray}
\label{COONECT}
D_iv_a=\partial_iv_a+f_{abc}A^b_iv_c,
\end{eqnarray}
\noindent
where $A^a_i$ is the Ashtekar connection.  The object $\overline{\nabla}_i$ is the unique torsion-free covariant derivative compatible with the spatial triad $e^a_i$, while $\nabla_i$ is the analogous covariant derivative when there is torsion present.

\section{Gauge curvature versus Riemannian curvature}

\noindent
To proceed with the goals of this paper we must first re-write the Ashtekar variables as spatial geometric variables.  Define an affine connection $\Gamma^k_{ij}$, in direct analogy to \cite{HIDDENSPATIAL}, such that
\begin{eqnarray}
\label{LETTER}
D_ie^a_j=\partial_ie^a_j+f^{abc}A^b_ie^c_j=\Gamma^k_{ij}e^a_k
\end{eqnarray}
\noindent
where $D_i$ is the gauge covariant derivative with respect to the $SO(3,C)$ gauge connection $A^a_i$.  Let us establish the effect of $D_i$ on symmetric and antisymmetric combinations of a spatial triad $e^a_i$, namely a densitized triad $\widetilde{\sigma}^i_a$ and a spatial 3-metric $h_{ij}$ given by
\begin{eqnarray}
\label{LETTER1}
\widetilde{\sigma}^i_a={1 \over 2}\epsilon^{ijk}\epsilon_{abc}e^b_je^c_k;~~h_{ij}=e^a_ie^a_j.
\end{eqnarray}
\noindent
Note also that $h_{ij}=(\hbox{det}\widetilde{\sigma})(\widetilde{\sigma}^{-1})^a_i(\widetilde{\sigma}^{-1})^a_j$, which is equivalent to (\ref{LETTER1}) for nondegenerate triads.\footnote{In the course of this paper we will show that $h_{ij}$ is the spatial part of a spacetime metric $g_{\mu\nu}$ solving the Einstein equations.}  The gauge covariant 
derivative acts on $\widetilde{\sigma}^i_a$ via
\begin{eqnarray}
\label{LETTER2}
D_m\widetilde{\sigma}^i_a=\epsilon^{ijk}\epsilon_{abc}e^b_jD_me^c_k=\epsilon^{ijk}\epsilon_{abc}e^b_j\Gamma^n_{mk}e^c_n,
\end{eqnarray}
\noindent
where we have used (\ref{LETTER}) and (\ref{LETTER1}).  Equation (\ref{LETTER2}) can then be re-written as
\begin{eqnarray}
\label{LETTER3}
D_m\widetilde{\sigma}^i_a=\epsilon^{ijk}\epsilon_{ljn}\Gamma^n_{mk}\widetilde{\sigma}^l_a
=\bigl(\delta^i_l\delta^k_n-\delta^i_n\delta^k_l\bigr)\Gamma^n_{mk}\widetilde{\sigma}^l_a=(\delta^i_l\Gamma^k_{mk}-\Gamma^i_{ml})\widetilde{\sigma}^l_a.
\end{eqnarray}
\noindent
Let us now impose the Ashtekar Gauss' law constraint on the densitized triad $D_i\widetilde{\sigma}^i_a=0$.  The trace of (\ref{LETTER3}) over spatial indices is given by
\begin{eqnarray}
\label{LETTER4}
D_i\widetilde{\sigma}^i_a=(\Gamma^k_{lk}-\Gamma^k_{kl})\widetilde{\sigma}^l_a=0.
\end{eqnarray}
\noindent
Therefore when Gauss' law (\ref{LETTER4}) holds, the trace of the torsion of $\Gamma^i_{jk}$ must vanish, where $T^i_{jk}=\Gamma^i_{[jk]}$ is the torsion.  Then in the decomposition
\begin{eqnarray}
\label{LETTER5}
T^i_{jk}=\epsilon_{jkm}S^{mi}+{1 \over 2}\bigl(\delta^i_ja_k-\delta^i_ka_j\bigr)
\end{eqnarray}
\noindent
where $S^{mi}=S^{im}$ is symmetric,\footnote{This is reminiscent of the decomposition of the structure constants of the Lie algebra for a Bianchi group \cite{WALD}, except we are referring to the full theory and not minisuperspace.} we have $T^i_{ik}=0$ which implies that $a_k=0$ and $T^i_{jk}=\epsilon_{jkm}S^{mi}$.  The effect of imposing Gauss' law would be to reduce the torsion $T^i_{jk}$ from nine to six degrees of freedom.\par
\indent
Having examined the consequences of the gauge covariant derivative for an antisymmetric combination of triads, let us now do so for a symmetric combination.  Acting on the 3-metric $h_{ij}$ we have
\begin{eqnarray}
\label{LETTER6}
D_mh_{ij}=\partial_mh_{ij}=D_m(e^a_ie^a_j),
\end{eqnarray}
\noindent
where we have used that the metric $h_{ij}$ is a gauge scalar due to the absence of internal indices.  Expanding (\ref{LETTER6}), we have
\begin{eqnarray}
\label{LETTER7}
\partial_mh_{ij}=e^a_i(D_me^a_j)+(D_me^a_i)e^a_j=e^a_i\Gamma^n_{mj}e^a_n+\Gamma^n_{mi}e^a_ne^a_j
\end{eqnarray}
\noindent
where we have used (\ref{LETTER}).  We can rewrite (\ref{LETTER7}) as
\begin{eqnarray}
\label{LETTER8}
\partial_mh_{ij}-\Gamma^n_{mj}h_{in}-\Gamma^n_{mi}h_{nj}=\nabla_mh_{ij}=0,
\end{eqnarray}
\noindent
which recognizes the covariant derivative of the 3-metric, seen as a second-rank tensor, with respect to the connection $\Gamma^k_{ij}$.  Equation (\ref{LETTER8}) states that the connection $\Gamma^i_{jk}$ is compatible with the 3-metric $h_{ij}$ constructed from the triads.  Note that this is not the 3-dimensional Levi--Civita connection $\Gamma^i_{(jk)}$ since $\Gamma^i_{jk}$ is in general allowed to have torsion.\footnote{Moreover, the torsion in general contains nine degrees of freedom, since there is nothing at this stage which requires Gauss' law to be satisfied as in (\ref{LETTER4}).}\par
\indent
We will now compute the curvature of the connection $\Gamma^i_{jk}$ using
\begin{eqnarray}
\label{LETTER9}
D_je^a_k=\Gamma_{jk}^me^a_m
\end{eqnarray}
\noindent
as in (\ref{LETTER}).  Acting on (\ref{LETTER9}) with a second gauge covariant derivative in the index $i$ and subtracting the result with $i$ and $j$ interchanged, we get
\begin{eqnarray}
\label{LETTER10}
[D_i,D_j]e^a_k=\Bigl(\partial_i\Gamma^n_{jk}-\partial_j\Gamma^n_{ik}+\Gamma^n_{im}\Gamma^m_{jk}-\Gamma^n_{jm}\Gamma^m_{ik}\Bigr)e^a_n=R^n_{kij}e^a_n.
\end{eqnarray}
\noindent
One recognizes in (\ref{LETTER10}) the 3-dimensional Riemann curvature tensor of the connection $\Gamma^i_{jk}$, which is a completely spatial tensor of fourth rank.  But the definition of the gauge covariant derivative allows us to re-write the left hand side of (\ref{LETTER10}) in terms of the $SO(3,C)$ gauge curvature
\begin{eqnarray}
\label{LETTER11}
[D_i,D_j]e^a_k=\epsilon_{ijl}\epsilon^{lmn}D_mD_ne^a_k=\epsilon_{ijl}f^{abc}B^l_be^c_k,
\end{eqnarray}
\noindent
where $B^i_a=\epsilon^{ijk}\partial_jA^a_k+{1 \over 2}\epsilon^{ijk}f_{abc}A^b_jA^c_k$ is the magnetic field for the connection $A^a_i$.  Equality of (\ref{LETTER11}) with (\ref{LETTER10}) implies that 
\begin{eqnarray}
\label{LETTER12}
\epsilon_{ijl}f^{abc}B^l_be^c_k=R^n_{kij}e^a_n=R_{nkij}E^n_a.
\end{eqnarray}
\noindent
In (\ref{LETTER12}) we have defined $E^n_a$ as the matrix inverse of the triad $e^a_i$, such that 
\begin{eqnarray}
\label{MATRIXINVERSE}
E^n_ae^a_m=\delta^n_m;~~E^n_ae^b_n=\delta_a^b.
\end{eqnarray}
\noindent
Transferring $E^n_a$ to the left hand side of (\ref{LETTER12}) and using (\ref{MATRIXINVERSE}), we have
\begin{eqnarray}
\label{LETTER13}
R_{nkij}=\epsilon_{ijl}f^{cab}e^c_ke^a_nB^l_b=\epsilon_{ijl}\epsilon_{knm}\widetilde{\sigma}^m_bB^l_b,
\end{eqnarray}
\noindent
where we have used (\ref{LETTER1}).  The result of this section has been to show that the $SO(3,C)$ gauge curvature encoded in the Ashtekar magnetic field $B^i_a$ is directly related to a completely spatial Riemann curvature tensor having no temporal components.  This is the curvature tensor of a 3-dimensional space with torsion, which we will refer to as $Q^{(3)}$.

\section{Ingredients for the Einstein--Hilbert action}

\noindent
In the next two sections we will show how Einstein's General Relativity of 4-dimensional spacetime follows from the intrinsic spatial 3-geometry of $Q^{(3)}$.  First expand the full Riemann curvature using 
the result of (\ref{LETTER10})
\begin{eqnarray}
\label{LETTER31}
R^n_{kij}(Q^{(3)})=\partial_i\Gamma^n_{jk}-\partial_j\Gamma^n_{ik}+\Gamma^n_{im}\Gamma^m_{jk}-\Gamma^n_{jm}\Gamma^m_{ik}.
\end{eqnarray}
\noindent
Then split the affine connection $\Gamma^i_{jk}$ into a part compatible with the 3-metric $h_{ij}$ and a part due to torsion
\begin{eqnarray}
\label{LETTER32}
\Gamma^i_{jk}=\Gamma^i_{(jk)}+T^i_{jk},
\end{eqnarray}
\noindent
where the curvature of the metric compatible part $\Gamma^i_{(jk)}$, namely the Levi--Civita connection due to symmetry in lower indices, is given by
\begin{eqnarray}
\label{LETTER33}
R^n_{kij}[h]=\partial_i\Gamma^n_{(jk)}-\partial_j\Gamma^n_{(ik)}+\Gamma^n_{(im)}\Gamma^m_{(jk)}-\Gamma^n_{(jm)}\Gamma^m_{(ik)}.
\end{eqnarray}
\noindent
Substituting (\ref{LETTER32}) into (\ref{LETTER31}) and using (\ref{LETTER33}), we have
\begin{eqnarray}
\label{LETTER34}
R^n_{kij}(Q^{(3)})=R^n_{kij}[h]+T^n_{im}T^m_{jk}-T^n_{jm}T^m_{ik}\nonumber\\
+\partial_iT^n_{jk}+\Gamma^n_{(im)}T^m_{jk}+\Gamma^m_{(jk)}T^n_{im}
-\partial_jT^n_{ik}-\Gamma^n_{(jm)}T^m_{ik}-\Gamma^m_{(ik)}T^n_{jm}.
\end{eqnarray}
\noindent
Next, contract (\ref{LETTER34}) by summing over $n=i$ to obtain the 3-dimensional Ricci tensor, in conjunction with using $T^i_{im}=0$ as follows from (\ref{LETTER4}).\footnote{This is at this stage a simplifying assumption needed to obtain General Relativity.  We will show later that this condition, which follows from the Gauss' law constraint, is actually required as a consistency condition on the space $Q^{(3)}$.}  Then the first line of the right hand side of (\ref{LETTER34}) reduces to
\begin{eqnarray}
\label{LETTER35}
R_{kj}[h]+T^i_{im}T^m_{jk}-T^i_{jm}T^m_{ik}=R_{kj}[h]-T^i_{jm}T^m_{ik}
\end{eqnarray}
\noindent
and the second line reduces to
\begin{eqnarray}
\label{LETTER36}
\partial_iT^i_{jk}+\Gamma^i_{(im)}T^m_{jk}-\Gamma^i_{(jm)}T^m_{ik}-\Gamma^m_{(ik)}T^i_{jm}=\overline{\nabla}_iT^i_{jk},
\end{eqnarray}
\noindent
where one recognizes $\overline{\nabla}_iT^i_{jk}$ as the covariant divergence of the torsion $T^i_{jk}$ with respect to the Levi--Civita connection $\Gamma^i_{(jk)}$.  We will next 
re-combine (\ref{LETTER36}) with (\ref{LETTER35}) and contract the sum with $h^{jk}$ to form the three dimensional curvature scalar $R$ of $Q^{(3)}$.  Note that this contraction annihilates (\ref{LETTER36}) due to antisymmetry of the torsion $T^i_{jk}=T^i_{[jk]}$, and we are left with
\begin{eqnarray}
\label{LETTER37}
R(Q^{(3)})=R[h]-h^{kj}T^n_{jm}T^m_{nk}
\end{eqnarray}
\noindent
as follows from (\ref{LETTER34}).  Recall the following decomposition from (\ref{LETTER5}), which when the Gauss' law constraint is satisfied simplifies to
\begin{eqnarray}
\label{LETTER38}
T^n_{jm}=\epsilon_{jml}S^{ln}.
\end{eqnarray}
\noindent
Substituting (\ref{LETTER38}) into (\ref{LETTER37}), we obtain the following expression for the term quadratic in torsion
\begin{eqnarray}
\label{LETTER39}
h^{kj}T^n_{jm}T^m_{nk}=h^{kj}\epsilon_{ksn}\epsilon_{jmr}S^{sm}S^{nr}\nonumber\\
=(\hbox{det}h)^{-1}\bigl(h_{nr}h_{sm}-h_{nm}h_{sr}\bigr)S^{rn}S^{sm}
=(\hbox{det}h)^{-1}\bigl((\hbox{tr}S)^2-S_{sm}S^{sm}\bigr).
\end{eqnarray}
\noindent
Let us make the definition
\begin{eqnarray}
\label{LETTER40}
S^{ij}=\beta\sqrt{h}K^{ij}
\end{eqnarray}
\noindent
where $\beta$ is a parameter which will be specified later.  Then substitution of (\ref{LETTER40}) into (\ref{LETTER39}) and (\ref{LETTER37}) yields
\begin{eqnarray}
\label{LETTER41}
R(Q^{(3)})=R[h]-\beta^2\bigl((\hbox{tr}K)^2-\hbox{tr}K^2\bigr).
\end{eqnarray}
\noindent
Multiplication of (\ref{LETTER41}) by $\sqrt{-g}=N\sqrt{h}$ and integration over spacetime yields
\begin{eqnarray}
\label{LETTER42}
I=\int{dt}\int_{\Sigma}d^3xN\sqrt{h}\Bigl({^{(3)}}R[h]-\beta^2\bigl((\hbox{tr}K)^2-\hbox{tr}K^2\bigr)\Bigr).
\end{eqnarray}
\noindent
If one could identify $K_{ij}$ with the extrinsic curvature of 3-space $\Sigma$, then the right hand side of (\ref{LETTER42}) for $\beta=i$ would be the Einstein--Hilbert action.  But this identification cannot be made arbitrarily.  So we need justification for the choice $a_k=0$ leading to (\ref{LETTER40}), as well as a physical principle for identifying $K_{ij}$ with extrinsic curvature.  We will provide the latter in the next section, and relegate the former to sections 5 and 6.

\section{The canonical structure}

Before the identification of (\ref{LETTER42}) with the Einstein--Hilbert action can be made, the appropriate canonical structure must be put in place.  Consider an infinitesimal $SO(3,C)$ gauge transformation and spacetime diffeomorphism $\delta_{\vec{\eta},\vec{\xi}}$ parameterized by $\vec{\eta}=\eta^a$ and $\xi=\xi^{\mu}$ respectively.  The spacetime metric $g_{\mu\nu}$ and a 4-dimensional $SO(3,C)$ gauge connection $A^a_{\mu}$ transform as \cite{BLAGOJEVIC}\footnote{In (\ref{LEEE}) we have used $\delta_{\vec{\eta}}g_{\mu\nu}=0$, namely that the spacetime metric is gauge-invariant.  Also, $L_{\xi}$ is the Lie derivative along the vector field $\xi^{\mu}$.}
\begin{eqnarray}
\label{LEEE}
\delta_{\vec{\eta},{\xi}}g_{\mu\nu}=L_{\xi}g_{\mu\nu}=\xi^{\sigma}\partial_{\sigma}g_{\mu\nu}+(\partial_{\mu}\xi^{\sigma})g_{\sigma\nu}+(\partial_{\nu}\xi^{\sigma})g_{\sigma\mu};\nonumber\\
\delta_{\vec{\eta},{\xi}}A^a_{\nu}=L_{\xi}A^a_{\nu}+\delta_{\vec{\eta}}A^a_{\nu}=D_{\nu}(\xi^{\mu}A^a_{\mu})+\xi^{\mu}F^a_{\mu\nu}-D_{\nu}\eta^a.
\end{eqnarray}
\noindent
We will show that the restriction of (\ref{LEEE}) to purely spatial fields $h_{ij}\equiv{g}_{ij}\subset{g}_{\mu\nu}$ and $A^a_i\subset{A}^a_{\mu}$ provides the ingredients needed for the canonical structure.  
For what follows we will restrict the spacetime diffeomorphisms to the form
\begin{eqnarray}
\label{CONSIDER}
\xi^{\nu}=\delta^{\nu}_0+\delta^{\nu}_kN^k,
\end{eqnarray}
\noindent
which consists of a spatial part $N^k$ and a temporal part $\delta^{\nu}_0$.  Substituting (\ref{CONSIDER}) into the spatial part of the second equation of (\ref{LEEE}), we obtain
\begin{eqnarray}
\label{LEEE3}
\delta_{\vec{\eta},\vec{\xi}}A^a_i
=D_i\bigl((\delta^{\mu}_0+N^k\delta^{\mu}_k)A^a_{\mu}\bigr)+(\delta^{\mu}_0+N^k\delta^{\mu}_k)F^a_{\mu{i}}\nonumber\\
=D_iA^a_0+D_i(N^kA^a_k)+F^a_{0i}+N^kF^a_{ki}-D_i\eta^a\nonumber\\
=\dot{A}^a_i+D_i(N^kA^a_k-\eta^a)+\epsilon_{ilk}B^l_aN^k,
\end{eqnarray}
\noindent
where we have used $F^a_{0i}=\dot{A}^a_i-D_iA^a_0$.  Defining $A^a_0\equiv\eta^a-N^kA^a_k$ as the temporal component of the connection, multiplying (\ref{LEEE3}) by $\widetilde{\sigma}^i_a$ and integrating over spacetime by parts we obtain the following integrand
\begin{eqnarray}
\label{LEEE4}
\widetilde{\sigma}^i_a\delta_{\vec{\eta},{\xi}}A^a_i=\widetilde{\sigma}^i_a\dot{A}^a_i+A^a_0D_i\widetilde{\sigma}^i_a+\epsilon_{ilk}\widetilde{\sigma}^i_aB^l_aN^k.
\end{eqnarray}
\noindent
Note that (\ref{LEEE4}) is the same as (\ref{YANGMILL2}) and (\ref{YANGMILL3}) with regard to $\widetilde{\sigma}^i_a\dot{A}^a_i$ and the Gauss' law and diffeomorphism constraint terms.  Having shown that the aforementioned prescription correctly produces the canonical structure and kinematic constraints (Gauss' law and diffeomorphism constraints) for the Ashtekar formulation, we will now follow suit for the metric case.\par
\indent
First we must substitute (\ref{CONSIDER}) into the spatial part of the first equation of (\ref{LEEE}) which gives
\begin{eqnarray}
\label{CONSIDER1}
\delta_{\vec{\eta},\vec{\xi}}g_{ij}=
(\delta^{\sigma}_0+\delta^{\sigma}_kN^k)\partial_{\sigma}g_{ij}+\partial_i(\delta^{\sigma}_kN^k)g_{\sigma{j}}+\partial_j(\delta^{\sigma}_kN^k)g_{\sigma{i}}\nonumber\\
=\partial_0g_{ij}+N^k\partial_kg_{ij}+(\partial_iN^k)g_{kj}+(\partial_jN^k)g_{ik},
\end{eqnarray}
\noindent
whence the transformations $\delta_{\vec{\eta}}$ act trivially.  To put (\ref{CONSIDER1}) into a more familiar form let us rewrite the partial deriatives in terms of covariant derivatives with respect to the Levi--Civita 
connection $\Gamma^k_{(ij)}$, using
\begin{eqnarray}
\label{CONSIDER2}
\partial_iN^k=\overline{\nabla}_iN^k-\Gamma^k_{(im)}N^m;~~\partial_kg_{ij}=\overline{\nabla}_kg_{ij}+\Gamma^m_{(ki)}g_{mj}+\Gamma^m_{(kj)}g_{mi}.
\end{eqnarray}
\noindent
Making the identification $g_{ij}=h_{ij}$ and substituting (\ref{CONSIDER2}) into (\ref{CONSIDER1}), we have
\begin{eqnarray}
\label{LEEE1}
\delta_{\vec{\eta},{\xi}}h_{ij}=\dot{h}_{ij}+N^k\partial_kh_{ij}+(\partial_iN^k)h_{kj}+(\partial_jN^k)h_{ki}\nonumber\\
=\dot{h}_{ij}+N^k\overline{\nabla}_kh_{ij}+N^k\Gamma^m_{ki}h_{mj}+N^k\Gamma^m_{(kj)}h_{im}\nonumber\\
+\bigl(\overline{\nabla}_iN^k-\Gamma^k_{(im)}N^m\bigr)h_{kj}+\bigl(\overline{\nabla}_jN^k-\Gamma^k_{(jm)}N^m\bigr)h_{ki}.
\end{eqnarray}
\noindent
The pure Levi--Civita connection terms in (\ref{LEEE1}) cancel out, which can be seen by a relabelling of indices.  Using this fact in conjunction with $\overline{\nabla}_kh_{ij}=0$ due to metric compatability, then (\ref{LEEE1}) simplifies to
\begin{eqnarray}
\label{CONSIDER3}
\delta_{\vec{\eta},{\xi}}h_{ij}=\dot{h}_{ij}+\overline{\nabla}_iN_j+\overline{\nabla}_jN_i.
\end{eqnarray}
\noindent
Multiplication of (\ref{CONSIDER3}) by $\pi^{ij}$ combined with an integration over spacetime and by parts yields the integrand
\begin{eqnarray}
\label{LEEE2}
\pi^{ij}\delta_{\vec{\eta},{\xi}}h_{ij}=\pi^{ij}\dot{h}_{ij}+2\pi^{ij}\overline{\nabla}_iN_j\longrightarrow\pi^{ij}\dot{h}_{ij}-2N_j\overline{\nabla}_i\pi^{ij}.
\end{eqnarray}
\noindent
In analogy with (\ref{LEEE4}) for the Ashtekar case, equation (\ref{LEEE2}) should be the canonical structure for metric General Relativity combined with a kinematic constraint, provided that $\pi^{ij}$ can be identified with the conjugate momentum for the 3-metric $h_{ij}$.  We will now make this association more precise.

\subsection{Legendre transformation to the Einstein--Hilbert action}

\noindent
To complete the demonstration that the intrinsic spatial 3-geometry of $Q^{(3)}$ includes 4-dimensional General Relativity, let us use the canonical structure $\pi^{ij}\dot{h}_{ij}$ to Legendre-transform (\ref{LETTER42}) into the Hamiltonian description, where
\begin{eqnarray}
\label{LETTER43}
\pi^{ij}=\beta\sqrt{h}\bigl(K^{ij}-h^{ij}(\hbox{tr}K)\bigr),
\end{eqnarray}
\noindent
is defined as the momentum canonically conjugate to the 3-metric $h_{ij}$ with $K_{ij}$ as given in (\ref{LETTER42}).  Inversion of (\ref{LETTER43}) yields
\begin{eqnarray}
\label{LETTER431}
K^{ij}={1 \over {\beta\sqrt{h}}}\bigl(\pi^{ij}-{1 \over 2}h^{ij}(\hbox{tr}\pi)\bigr),
\end{eqnarray}
\noindent
then substitution of (\ref{LETTER431}) into (\ref{LETTER42}) yields
\begin{eqnarray}
\label{LETTER432}
I=\int{dt}\int_{\Sigma}d^3xN\Bigl[\sqrt{h}{^{(3)}R[h]}+{1 \over {\sqrt{h}}}\Bigl(\pi^{ij}\pi_{ij}-{1 \over 2}(\hbox{tr}\pi)^2\Bigr)\Bigr]
\end{eqnarray}
\noindent
whence the parameter $\beta$ has cancelled out.  To complete the Legendre transformation of (\ref{LETTER432}), we need to express $K_{ij}$ in terms of $\dot{h}_{ij}$.  This requires that we make the identification
\begin{eqnarray}
\label{LETTER433}
\dot{h}_{ij}=2\beta{N}K_{ij}+\overline{\nabla}_iN_j+\overline{\nabla}_jN_i,
\end{eqnarray}
\noindent
where $N$ is the lapse function and $N_i$ the shift vector.  Note that (\ref{LETTER433}) is the statement that $K_{ij}$ is essentially the Lie derivative 
of $h_{ij}$ in the direction of the timelike 4-vector $\xi^{\mu}=\delta^{\mu}_0$.  Note that $2\beta{N}K_{ij}=\underline{N}\bigl(2\pi_{ij}-h_{ij}(\hbox{tr}\pi)\bigr)$, which follows from (\ref{LETTER431}).  The canonical one form $\pi^{ij}\delta{h}_{ij}$ implies that
\begin{eqnarray}
\label{LETTER434}
\pi^{ij}\dot{h}_{ij}={N \over {\sqrt{h}}}\bigl(2\pi_{ij}\pi^{ij}-(\hbox{tr}\pi)^2\bigr)+2\pi^{ij}\overline{\nabla}_iN_j,
\end{eqnarray}
\noindent
where we have used the symmetry of $\pi^{ij}$.  Then the Legendre transformation of (\ref{LETTER432}) is given by
\begin{eqnarray}
\label{LETTER435}
\boldsymbol{H}=\int_{\Sigma}d^3x\pi^{ij}\dot{h}_{ij}-I
=\int_{\Sigma}d^3x\Bigl(2\pi^{ij}\overline{\nabla}_iN_j\nonumber\\
+N\Bigl(-\sqrt{h}{^{(3)}R[h]}+{1 \over {\sqrt{h}}}\bigl(\pi_{ij}\pi^{ij}-{1 \over 2}(\hbox{tr}\pi)^2\bigr)\Bigr).
\end{eqnarray}
\noindent
Integrating by parts and discarding boundary terms, (\ref{LETTER435}) becomes
\begin{eqnarray}
\label{LETTER436}
\boldsymbol{H}=\int_{\Sigma}d^3x(N^iH_i+NH)
\end{eqnarray}
\noindent
where $H_i$ and $H$ are the Hamiltonian and diffeomorphism constraints on the full Einstein--Hilbert metric phase space $\Omega_{EH}=(h_{ij},\pi^{ij})$, given by
\begin{eqnarray}
\label{METRICCONN}
H=\pi^{ij}\pi_{ij}-{1 \over 2}(\hbox{tr}\pi)^{2}-\sqrt{h}{^{(3)}R[h]}=0;~~H_i=\overline{\nabla}^j\pi_{ij}=0.
\end{eqnarray}
\noindent
Therefore when one makes the identifications (\ref{LETTER4}), (\ref{LETTER40}) and (\ref{LETTER433}), 
then ${Q}^{(3)}$ yields the Einstein--Hilbert action as inherent in its 3+1 decomposition.\footnote{This provides additional justification for the prescription leading to (\ref{LEEE2}).}  Note that the 
Hamiltonian (\ref{LETTER435}) is insensitive to the presence of the parameter $\beta$, but the action (\ref{LETTER42}) is not.  Equation (\ref{LETTER43}) implies that for $\beta=\pm{i}$, one is in a tunneling configuration in the quantum theory since the momentum $\pi^{ij}$ is imaginary.  For $\beta=\pm{1}$ the theory is in an oscillatory configuration since $\pi^{ij}$ is real.  This also suggests the identification of $\beta$ with the Immirzi parameter of the Ashtekar formalism \cite{IMMIRZI}.

\section{Yang--Mills spatial geometry of $Q^{(3)}$}

We have shown how the intrinsic spatial geometry of a 3-dimensional space $Q^{(3)}$ with torsion leads via Gauss' law to the 4-dimensional Einstein--Hilbert action.  Using $Q^{(3)}$, we will now show how General Relativity can be thought of as a sort of `generalized' Yang--Mills theory.  In the developments of \cite{HIDDENSPATIAL} and \cite{HIDDENSPATIAL1}, the magnetic 
field $B^i_a$ or a densitized version plays the role of the triad $E^i_a$.  This enables one to rewrite (\ref{LETTER13}) completely in terms of a metric $\phi^{ij}=B^i_bB^j_b$ constructed from the magnetic field $B^i_a$, thus leading to the Einstein space condition (\ref{YANGMILL5}).  But we would like to extend this concept to more general spacetime geometries.  Let us now introduce the Ansatz 
\begin{eqnarray}
\label{LETTER14}
\widetilde{\sigma}^k_b=\beta\Psi_{bf}B^k_f;~(\hbox{det}\Psi)\neq{0},
\end{eqnarray}
\noindent
where $\Psi_{bf}\in{SO}(3,C)\otimes{SO}(3,C)$ and $\beta$ is a numerical constant which remains to be determined.  Substituting (\ref{LETTER14}) into (\ref{LETTER13}), we obtain
\begin{eqnarray}
\label{LETTER15}
R_{ijmn}={1 \over \beta}\epsilon_{ijl}\epsilon_{mnk}\widetilde{\sigma}^l_f\widetilde{\sigma}^k_b\Psi^{-1}_{bf}.
\end{eqnarray}
\noindent
First note on account of (\ref{LETTER14}) that (\ref{LETTER15}) can be written in the equivalent form
\begin{eqnarray}
\label{LETTER44}
R_{ijmn}={1 \over \beta}\epsilon_{ijl}\epsilon_{mnk}\widetilde{\sigma}^l_f\widetilde{\sigma}^k_b\Psi^{-1}_{bf}
=\beta\epsilon_{ijl}\epsilon_{mnk}B^l_fB^k_b\Psi_{bf}.
\end{eqnarray}
\noindent
By taking in the average of both forms in (\ref{LETTER44}) we can write the Riemann curvature tensor of $Q^{(3)}$ as
\begin{eqnarray}
\label{LETTER45}
R_{ijmn}=\epsilon_{ijl}\epsilon_{mnk}\Bigl[{1 \over {\beta^2}}\widetilde{\sigma}^l_f\widetilde{\sigma}^k_b(\Psi^{-1})^{bf}+\Psi_{bf}B^l_bB^k_f\Bigr]\nonumber\\
=-\beta\epsilon_{ijl}\epsilon_{mnk}T^{lk}+{1 \over 2}\beta\Bigl(1+{1 \over {\beta^2}}\Bigr)\epsilon_{ijl}\epsilon_{mnk}\widetilde{\sigma}^l_b\widetilde{\sigma}^k_f(\Psi^{-1})^{bf}
\end{eqnarray}
\noindent
where we have defined
\begin{eqnarray}
\label{HAVEITT}
T^{ij}={1 \over 2}\Bigl[\widetilde{\sigma}^l_f\widetilde{\sigma}^k_b(\Psi^{-1})^{bf}-\Psi_{bf}B^l_bB^k_f\Bigr].
\end{eqnarray}
\noindent
For the choice $\beta=\pm{i}$ the second term on the right hand side of (\ref{LETTER45}) vanishes.  Then double contraction yields the 3-dimensional curvature scalar of $Q^{(3)}$
\begin{eqnarray}
\label{LETTER46}
R=h^{im}h^{jn}R_{ijmn}=(\hbox{det}h)^{-1}h_{lk}T^{lk},
\end{eqnarray}
\noindent
where we have used the property of determinants of three by three matrices.  To obtain the Einstein--Hilbert action as inferred from (\ref{LETTER42}), we must multiply (\ref{LETTER46}) by $\sqrt{-g}=N\sqrt{h}$ and integrate over 
spacetime.  Defining $\underline{N}=N/\sqrt{h}=N(\hbox{det}\widetilde{\sigma})^{-1/2}$ this yields
\begin{eqnarray}
\label{LETTER47}
I_{EH}=\int_md^4x\sqrt{-g}{^{(4)}R}=\int{dt}\int_{\Sigma}d^3x\underline{N}h_{lk}T^{lk}.
\end{eqnarray}
\noindent
The left hand side of (\ref{LETTER47}) is the Einstein--Hilbert action both for 4-dimensional spacetime and for $Q^{(3)}$.  The right hand side is the action for Yang--Mills theory coupled to the spacetime metric $g_{\mu\nu}$, where $\Psi_{bf}$ replaces the Cartan--Killing form for $SO(3)$, and plays a dual role as the Yang--Mills coupling constant.  It follows, on account of (\ref{LETTER14}), that the Yang--Mills field is also the gravitational field of the theory, which in a way makes this a 
self-coupling.\footnote{This observation will lead us to the instanton representation of gravity, which will be introduced later in this paper.}  Moreover, $h_{ij}$ as defined by (\ref{YANGMILLS121}) is the spatial part of the spacetime metric $g_{\mu\nu}$ solving the equations of motion for $I_{EH}$.\par
\indent
 
\subsection{Relation with the gravitational degrees of freedom}

We will now take one step back to explore the precise relation of $\Psi_{bf}$, as defined by (\ref{LETTER14}), to the space $Q^{(3)}$.  The 3-dimensional Ricci tensor for $Q^{(3)}$, is obtained by contraction of (\ref{LETTER15}) with $h^{jn}$
\begin{eqnarray}
\label{LETTER16}
R_{im}=h^{jn}R_{ijmn}={1 \over \beta}h^{jn}\epsilon_{jli}\epsilon_{mnk}\widetilde{\sigma}^l_f\widetilde{\sigma}^k_b\Psi^{-1}_{bf}\nonumber\\
={1 \over \beta}(\hbox{det}h)^{-1}\bigl(h_{lm}h_{ik}-h_{lk}h_{im}\bigr)\widetilde{\sigma}^l_f\widetilde{\sigma}^k_b\Psi^{-1}_{bf}
={1 \over \beta}\bigl(e^b_ie^f_m-h_{im}e^{bk}e^f_k\bigr)\Psi^{-1}_{bf}.
\end{eqnarray}
\noindent
Another contraction of (\ref{LETTER16}) with $h^{im}$ will yield the three dimensional curvature scalar
\begin{eqnarray}
\label{LETTER17}
R=h^{im}R_{im}=-{2 \over \beta}(e^{bk}e^f_k)\Psi^{-1}_{bf}.
\end{eqnarray}
\noindent
From (\ref{LETTER17}) and (\ref{LETTER16}) we can form the three dimensional Einstein tensor
\begin{eqnarray}
\label{LETTER18}
G_{im}=R_{im}-{1 \over 2}h_{im}R={1 \over \beta}e^b_ie^f_m\Psi^{-1}_{bf}.
\end{eqnarray}
\noindent
One sees from (\ref{LETTER18}) that the inverse matrix $\Psi^{-1}_{bf}$ has the physical interpretation of the Einstein tensor for a three dimensional space $Q^{(3)}$ with torsion, expressed in the triad frame.  Let us perform the following decomposition
\begin{eqnarray}
\label{LETTER19}
\Psi^{-1}_{bf}=\delta_{bf}\varphi+\psi_{bf}+\epsilon_{bfd}\psi^d
\end{eqnarray}
\noindent
where $\psi_{bf}$ is symmetric and traceless.  For $\psi^d=0$ and $\varphi=-{\Lambda \over 3}$, where $\Lambda$ is the cosmological constant, we have
\begin{eqnarray}
\label{LETTER20}
\Psi^{-1}_{bf}=-\Bigl({\Lambda \over 3}\Bigr)\delta_{bf}+\psi_{bf},
\end{eqnarray}
\noindent
whence $\psi_{bf}$ takes on the interpretation of the spinorial form of the self-dual part of the Weyl curvature tensor as introduced in \cite{PENROSERIND}, \cite{MACCALLUM}.  While we have already explicitly obtained 4-dimensional General Relativity from $Q^{(3)}$, this might appear naively to present a paradox as it is commonly stated that the Weyl tensor $C_{ijkl}$ should vanish in 3-dimensions.  Indeed, from the decomposition of the Riemann tensor 
\begin{eqnarray}
\label{3DRIEMANN}
^{(3)}R_{ijkl}=h_{ik}R_{jl}-h_{il}R_{jk}+h_{jl}R_{ik}-h_{jk}R_{il}-{1 \over 2}R\bigl(h_{ik}h_{jl}-h_{il}h_{jk}\bigr),
\end{eqnarray} 
\noindent
one sees that $C_{ijkl}$ does not explicitly appear.  However, substitution of the left hand sided equality of (\ref{LETTER18}) into (\ref{3DRIEMANN}) yields
\begin{eqnarray}
\label{3DRIEMANN1}
^{(3)}R_{ijkl}=h_{ik}G_{jl}-h_{il}G_{jk}+h_{jl}G_{ik}-h_{jk}G_{il}.
\end{eqnarray}
\noindent
It is not difficult to see that substitution of the right hand sided equality of (\ref{LETTER18}) into (\ref{3DRIEMANN}) yields (\ref{LETTER15}) which as we have shown is really the full 4-dimensional Riemann curvature tensor.  This suggests that the Weyl tensor in (\ref{3DRIEMANN}) is not really zero, but rather is contained in $\Psi^{-1}_{bf}$.  Therefore, (\ref{LETTER47}) suggests that General Relativity can be thought of literally as a `generalized' Yang--Mills theory with the Weyl tensor replacing the Yang--Mills coupling constant.  We will prove more rigorously that this is indeed the case.
  
\section{The Bianchi identities for $Q^{(3)}$}

\noindent
Having related 4-dimensional metric General Relativity with Yang--Mills theory via $Q^{(3)}$, the Ashtekar variables and the Ansatz (\ref{LETTER14}), we will now establish the nature of the coupling field $\Psi_{bf}$ by exploiting a few properties 
of $Q^{(3)}$.  Since $G_{ij}$ is an Einstein tensor then it should satisfy the contracted Bianchi identities for $Q^{(3)}$.  The first Bianchi identity is $R_{i[jmn]}=0$, and using (\ref{LETTER15}) we have
\begin{eqnarray}
\label{BIANCHII}
R_{ijmn}\epsilon^{ijm}=\epsilon^{ijm}\epsilon_{jli}\epsilon_{mnk}\widetilde{\sigma}^l_f\widetilde{\sigma}^k_b\Psi^{-1}_{bf}\nonumber\\
=2\epsilon_{kmn}\widetilde{\sigma}^k_b\widetilde{\sigma}^m_f\Psi^{-1}_{bf}=2(\hbox{det}\widetilde{\sigma})(\widetilde{\sigma}^{-1})^d_n\epsilon_{dbf}\Psi^{-1}_{bf}=0
\end{eqnarray}
\noindent
where we have used $(\hbox{det}\widetilde{\sigma})\neq{0}$.  The result is that the first Bianchi identity for $Q^{(3)}$ requires that the antisymmetric part of $\Psi_{bf}$ be zero, or that $\Psi_{bf}=\Psi_{bf}$ be symmetric.\par
\indent
Let us impose the following conditions on on $G_{ij}$
\begin{eqnarray}
\label{LETTER22}
\epsilon^{kij}G_{ij}=0;~~\Lambda+h^{im}G_{im}=0,
\end{eqnarray}
\noindent
which from (\ref{LETTER17}) implies the Einstein space condition $R=2\Lambda$ with $k=6\Lambda$ in (\ref{YANGMILL5}).  Equations (\ref{LETTER22}) imply the following four constraints on the nine components of $\Psi_{bf}$
\begin{eqnarray}
\label{LETTER21}
\epsilon_{dbf}\Psi^{-1}_{bf}=0;~~\beta\Lambda+\hbox{tr}\Psi^{-1}=0.
\end{eqnarray}
\noindent
But 4-dimensional General Relativity should have two unconstrained degrees of freedom per point, which implies that there must be three additional constraints on the five remaining components of $\Psi_{bf}$.  To determine these constraints, it will be instructive to examine the associated constraints on $G_{ij}$.  Since $G_{ij}$ is an Einstein tensor, then it should also satisfy the contracted second Bianchi identity for $Q^{(3)}$, $\nabla^jG_{ij}=0$ in addition to (\ref{LETTER22}).  Let us act with the gauge covariant derivative $D_k$ on the quantity 
\begin{eqnarray}
\label{LETTER23}
G_{ij}=e^b_ie^f_j\Psi^{-1}_{bf}.
\end{eqnarray}
\noindent
Since $G_{ij}$ does not have internal indices, then its gauge covariant derivative is the same as its partial derivative.  Hence, acting with the gauge covariant derivative on (\ref{LETTER23}), we have for the left hand side that 
$D_kG_{ij}=\partial_kG_{ij}$.  Expanding the right hand side and using (\ref{LETTER}), we have
\begin{eqnarray}
\label{LETTER24}
D_kG_{ij}=\partial_kG_{ij}=(D_ke^b_i)e^f_j\Psi^{-1}_{bf}+e^b_i(D_ke^f_j)\Psi^{-1}_{bf}+e^b_ie^f_j(D_k\Psi^{-1}_{bf})\nonumber\\
=\Gamma^m_{ki}e^b_me^f_j\Psi^{-1}_{bf}+e^b_i\Gamma^m_{kj}e^f_m\Psi^{-1}_{bf}+e^b_ie^f_j(D_k\Psi^{-1}_{bf})\nonumber\\
=\Gamma^m_{ki}G_{mj}+\Gamma^m_{kj}G_{im}+e^b_ie^f_j(D_k\Psi^{-1}_{bf}).
\end{eqnarray}
\noindent
Transferring the $\Gamma^i_{jk}$ terms to the left hand side, equation (\ref{LETTER24}) can be rewritten as
\begin{eqnarray}
\label{LETTER25}
\nabla_kG_{ij}=e^b_ie^f_j(D_k\Psi^{-1}_{bf}),
\end{eqnarray}
\noindent
whence one recognizes the definition of the covariant derivative of $G_{ij}$, seen as a tensor of second rank, with respect to the connection $\Gamma^k_{ij}$.  The right hand side of (\ref{LETTER25}) will have a part due 
to $\partial_k\Psi^{-1}_{bf}$ and a part free of spatial derivatives.  For the first part we will use the matrix identity
\begin{eqnarray}
\label{LETTER251}
\partial_k\Psi^{-1}_{bf}=-\Psi^{-1}_{ba}(\partial_k\Psi_{ac})\Psi^{-1}_{cf}.
\end{eqnarray}
\noindent
Hence, expanding (\ref{LETTER25}) while using (\ref{LETTER251}) and index reshuffling yields
\begin{eqnarray}
\label{LETTER26}
e^b_ie^f_j(D_k\Psi^{-1}_{bf})=e^b_ie^f_j\Bigl(-\Psi^{-1}_{ba}(\partial_k\Psi_{ad})\Psi^{-1}_{df}-f_{bcd}A^c_k\Psi^{-1}_{df}-f_{fcd}A^c_k\Psi^{-1}_{bd}\Bigr)\nonumber\\
=-e^b_ie^f_j\Psi^{-1}_{ba}\Bigl(\partial_k\Psi_{ad}+\Psi_{ag}f_{gcd}A^c_k+f_{gca}A^c_k\Psi_{gd}\Bigr)\Psi^{-1}_{df}.
\end{eqnarray}
\noindent
We have used the definition of the gauge covariant derivative of a second rank $SO(3,C)$ tensor in (\ref{LETTER26}).  Note that for the special case $\Psi^{-1}_{ab}=\delta_{ab}k$ for numerically constant $k$ causes (\ref{LETTER26}) to vanish, which yields the Einstein space in \cite{HIDDENSPATIAL}.  In the general case $\Psi_{bf}$ contains gravitational degrees of freedom, which implies more general geometries.\par
\indent
To form the contracted second contracted Bianchi identity for the Einstein tensor $G_{ij}$, contract (\ref{LETTER25}) with $h^{jk}$, which yields
\begin{eqnarray}
\label{LETTER27}
\nabla^jG_{ij}=-\Psi^{-1}_{ba}\Psi^{-1}_{df}e^b_iE^j_fD_j\Psi_{ad}=-(\Psi^{-1}_{ba}e^b_j)(\Psi^{-1}_{df}E^j_f)D_j\Psi_{ad}.
\end{eqnarray}
\noindent
Using $E^i_a=(\hbox{det}\widetilde{\sigma})^{-1/2}\widetilde{\sigma}^i_a$ in conjunction with (\ref{LETTER14}), then (\ref{LETTER27}) reduces to
\begin{eqnarray}
\label{LETTER28}
\nabla^jG_{ij}=-(G_{jm}E^m_a)(\hbox{det}\widetilde{\sigma})^{-1/2}B^j_dD_j\Psi_{ad}.
\end{eqnarray}
\noindent
Defining $B^j_dD_j\Psi_{ad}\equiv\textbf{w}_d\{\Psi_{ad}\}$, then the Bianchi identity for $Q^{(3)}$ reduces to
\begin{eqnarray}
\label{LETTER29}
\nabla^jG_{ij}=-(\hbox{det}\widetilde{\sigma})^{-1/2}(G_{jm}E^m_a)\textbf{w}_d\{\Psi_{ad}\}=0.
\end{eqnarray}
\noindent
Hence if we require that $\textbf{w}_d\{\Psi_{ad}\}=0$, then this guarantees that the Bianchi identity is satisfied.  So augmenting the list of constraints (\ref{LETTER21}) and (\ref{LETTER22}) to
\begin{eqnarray}
\label{LETTER30}
\epsilon_{dbf}\Psi^{-1}_{bf}=0;~~\beta\Lambda+\hbox{tr}\Psi^{-1}=0;~~\textbf{w}_e\{\Psi_{ae}\}=0;\nonumber\\
\longrightarrow\epsilon^{kij}G_{ij}=0;~~\Lambda+h^{im}G_{im}=0;~~\nabla^jG_{ij}=0
\end{eqnarray}
\noindent
completes the list of constraints on our system in order that it exhibit two unconstrained degrees of freedom.  The first and second Bianchi identities for the space $Q^{(3)}$ imply that the traceless part of $\Psi^{-1}_{bf}$ is the self-dual Weyl tensor.  The trace of $\Psi^{-1}_{bf}$ is fixed by the requirement that the scalar curvature of $Q^{(3)}$ is twice the cosmological constant.  Equations (\ref{LETTER30}) generalize the results of \cite{HIDDENSPATIAL} to include more general geometries.\par
\indent

\section{Conclusion and discussion}

\noindent
This paper has elucidated on the relations amongst the Ashtekar variables, 3-dimensional intrinsic spatial geometry, 4-dimensional General Relativity, and Yang--Mills theory.  We have shown that the Gauss' law constraint in Ashtekar variables implies the existence of a 3-dimensional Riemannian space with torsion, defined as $Q^{(3)}$.  When Gauss' law holds, the torsion of $Q^{(3)}$ must contain six rather than nine degrees of freedom in order to yield a metric theory.  By associating these degrees of freedom with the extrinsic curvature tensor $K_{ij}$, we showed that the (3-dimensional) Riemann curvature tensor of $Q^{(3)}$ is the same as the (four dimensional) Riemann curvature tensor via the 3+1 ADM decomposition of General Relativity.  We have shown that the associated Einstein--Hilbert action is equivalent to an action 
defined on $Q^{(3)}$ corresponding to a Yang--Mills theory with a field $\Psi_{ae}$ playing the role of a coupling constant.  The field $\Psi_{ae}$ encodes the gravitational degrees of freedom of a 4-dimensional spacetime solving the Einstein equations.  We showed this via the Bianchi identities of the space $Q^{(3)}$, which proved that the traceless part of $\Psi^{-1}_{ae}$ is the self-dual Weyl curvature tensor exhibiting the two degrees of freedom of 4-dimensional General Relativity.\par
\indent

\subsection{The instanton representation of Plebanski gravity}

The main result of this paper has been to establish the physical foundations for the space $Q^{(3)}$ in relation to 4-dimensional General Relativity and Yang--Mills theory.  We will show that the first line of (\ref{LETTER30}) can be obtained from an action principle derivable from the Ashtekar variables, which implies the existence of a General Relativity formulation in terms of the field $\Psi_{ae}$.  Using the definition of the determinant of 3 by 3 matrices 
\begin{eqnarray}
\label{THEDEFINITION}
\hbox{det}\widetilde{\sigma}={1 \over 6}\epsilon_{ijk}\epsilon^{abc}\widetilde{\sigma}^i_a\widetilde{\sigma}^j_b\widetilde{\sigma}^k_c;
~~(\widetilde{\sigma}^{-1})^c_k={1 \over 2}\epsilon_{ijk}\epsilon^{abc}\widetilde{\sigma}^i_a\widetilde{\sigma}^j_b,
\end{eqnarray}
\noindent
then the action (\ref{YANGMILL2}) in Ashtekar variables can be written as
\begin{eqnarray}
\label{THEDEFINITION1}
I_{Ash}=\int{dt}\int_{\Sigma}d^3x\widetilde{\sigma}^i_a\dot{A}^a_i+A^a_0D_i\widetilde{\sigma}^i_a+\epsilon_{ijk}N^i\widetilde{\sigma}^j_aB^k_a-iN\sqrt{\hbox{det}\widetilde{\sigma}}\bigl(\Lambda+(\widetilde{\sigma}^{-1})^a_iB^i_a\bigr).
\end{eqnarray}
\noindent
Substitution of (\ref{LETTER14}) into (\ref{THEDEFINITION1}) yields the action
\begin{eqnarray}
\label{THEDEFINITION2}
I_{Inst}=\beta\int{dt}\int_{\Sigma}d^3x\Bigl[\Psi_{ae}B^i_e\dot{A}^a_i+A^a_0\textbf{w}_e\{\Psi_{ae}\}\nonumber\\
+\epsilon_{ijk}N^iB^j_aB^k_e\Psi_{ae}-i\beta^{-1/2}N\sqrt{\hbox{det}B}\sqrt{\hbox{det}\Psi}\bigl(\beta\Lambda+\hbox{tr}\Psi^{-1}\bigr)\Bigr],
\end{eqnarray}
\noindent
whose constraints are the first line of (\ref{LETTER30}).  We will refer to (\ref{THEDEFINITION2}) as the instanton representation of Plebanski gravity.  A few remarks are in order. (i) The phase space variables for $I_{Inst}$ are the field $\Psi_{ae}$ and the Ashtekar connection $A^a_i$. (ii) The connection $A^a_i$ can be seen as the spatial part of a 4-dimensional gauge connection $A^a_{\mu}$. (iii) The temporal part of $A^a_{\mu}$, given by $A^a_0$ smears the Gauss' law constraint of (\ref{THEDEFINITION2}).  But we have
\begin{eqnarray}
\label{BUTWEHAVE}
G_a=D_i\widetilde{\sigma}^i_a=\textbf{w}_e\{\Psi_{ae}\}\sim\nabla^iG_{ij}\sim{T}^i_{ik},
\end{eqnarray}
\noindent
which shows that the Gauss' law constraint $G_a$ is nothing other than the second Bianchi identity of $Q^{(3)}$.  Recall that Gauss' law was required in order for 4-dimensional metric General Relativity to follow from $Q^{(3)}$.  So the physical principle which demands that Gauss' law be satisfied is simply a consistency condition implied by the Bianchi identity.  This closes the missing link in the equivalence of intrinsic 3-dimensional spatial geometry to 4-dimensional General Relativity and to Yang--Mills theory.  Therefore, $I_{Inst}$ is another formulation of Einstein's General Relativity using new variables $\Psi_{ae}$ and $A^a_{\mu}$.  In Appendix A we will examine the effect of relaxing the Gauss' law constraint.\par
\indent
The action (\ref{THEDEFINITION2}) has been obtained from more fundamental principles in \cite{EYOITA}.  The future directions of research will include the establishment of $I_{Inst}$ as a new reformulation of general relativity and to fully explore its consequences at the classical and the quantum levels.  The action (\ref{THEDEFINITION2}) is related to an action appearing in \cite{SPINCON} which was used to obtain the pure spin connection formulation of General Relativity by Capovilla, Jacobson and Dell.  It is explained in \cite{EYOITA} the similarities and the differences between these actions.

\section{Appendix A: Relaxation of the metricity condition}

We have shown, when Gauss' law holds, that the space $Q^{(3)}$ is consistent with Einstein's 4-dimensional General Relativity.  In this appendix we will examine the situation when one chooses to relax Gauss' law.  Recall that the 3-dimensional Riemannian curvature of $Q^{(3)}$ is given by
\begin{eqnarray}
\label{IMPLIES2}
R^n_{ijm}={^{(3)}}R^n_{ijm}[h]+T^k_{jm}T^n_{ik}-T^k_{im}T^n_{jk}+\nabla_{[i}T^n_{j]m}.
\end{eqnarray}
\noindent
First contract $n$ with $j$ to form the 3-diemsional Ricci tensor
\begin{eqnarray}
\label{IMPLIES3}
R_{im}=R^j_{ijm}={^{(3)}}R_{im}[h]+T^k_{jm}T^j_{ik}-T^k_{im}T^j_{jk}+\nabla_{[i}T^j_{j]m}.
\end{eqnarray}
\noindent
Then contract (\ref{IMPLIES3}) with $h^{im}$ to form the 3-dimensional Ricci tensor.  This yields
\begin{eqnarray}
\label{IMPLIES4}
R=h^{im}R_{im}={^{(3)}}R[h]+h^{im}T^k_{jm}T^j_{ik}+h^{im}\nabla_{[i}T^j_{j]m},
\end{eqnarray}
\noindent
which eliminates one of the torsion squared terms due to antisymmetry.  The remaining torsion squared term is of the form $T^k_{jm}T_k^{jm}$, which suggests identification with an extrinsic curvature term squared.  Perform the following decomposition
\begin{eqnarray}
\label{DECOMPOSE}
T^n_{im}=\epsilon_{iml}K^{ln}+\delta^n_ia_m-\delta^n_ma_i,
\end{eqnarray}
\noindent
where $K^{ln}$ is symmetric.  The physical interpretation of $a_m$ arises from taking the trace of (\ref{DECOMPOSE}), which yields $a_m={1 \over 2}T^i_{im}$.  Substitution of (\ref{DECOMPOSE}) into the middle term of (\ref{IMPLIES4}) yields 
\begin{eqnarray}
\label{DECOMPOSE1}
h^{im}T^k_{jm}T^j_{ik}=h^{im}\epsilon_{ikn}\epsilon_{mlj}K^{kl}K^{nj}-2h^{ij}a_ia_j.
\end{eqnarray}
\noindent
Upon making the identification
\begin{eqnarray}
\label{DECOMPOSE2}
h^{im}\epsilon_{ikn}\epsilon_{mlj}=h_{kl}h_{nj}-h_{kj}h_{nl},
\end{eqnarray}
\noindent
The first term of (\ref{DECOMPOSE1}) indeed reduces to $(\hbox{tr}K)^2-\hbox{tr}K^2$, which confirms the interpretation of $K^{ij}$ as the extrinsic curvature tensor.  This also implies that $\epsilon_{ijk}$ is a tensor density, corresponding to curved space.  Substituting these results into (\ref{IMPLIES4}) yields
\begin{eqnarray}
\label{DECOMPOSE3}
R={^{(3)}}R+(\hbox{tr}K)^2-K^{ij}K_{ij}+\nabla^ma_m-2a^ma_m.
\end{eqnarray}
\noindent
The 3-vector $a_m$ is directly proportional to the nonmetricity of the theory.  For $a_m=0$ the theory becomes metric, whereupon the right hand side becomes the 3+1 decomposition of the four dimensional Riemann curvature tensor.  In the general case the decomposition of its three dimensional affine connection is given by
\begin{eqnarray}
\label{DECOMPOSE4}
\Gamma^k_{ij}=\Gamma^k_{(ij)}+\epsilon_{ijm}K^{mk}+\bigl(\delta^k_i\delta^n_j-\delta^k_j\delta^i_n\bigr)(\widetilde{\sigma}^{-1})^a_nG_a.
\end{eqnarray}
\noindent
One could also interpret the nonmetric theory (\ref{DECOMPOSE3}) as metric gravity coupled to a new field $a^m$.

\end{document}